\documentclass[aps,prl,twocolumn,showpacs,amssymb,groupedaddress]{revtex4}
\usepackage{graphicx}

\begin{document}


\title{Friction of a slider on a granular layer: Non-monotonic
  thickness dependence and effect of boundary conditions
}
\author{Saloome Siavoshi, Ashish V. Orpe and Arshad Kudrolli}
\affiliation{Department of Physics, Clark University, Worcester,
  Massachusetts 01610} 
\date{\today}
\begin{abstract}
  We investigate the effective friction encountered by a mass sliding
  on a granular layer as a function of bed thickness and boundary
  roughness conditions. The observed friction has minima for a small
  number of layers before it increases and saturates to a value which
  depends on the roughness of the sliding surface. We use an
  index-matched interstitial liquid to probe the internal motion of
  the grains with fluorescence imaging in a regime where the liquid
  has no significant effect on the measured friction. The shear profiles obtained
  as a function of depth show decrease in slip near the sliding
  surface as the layer thickness is increased. We propose that the
  friction depends on the degree of grain confinement relative to the
  sliding surfaces.
\end{abstract}




\maketitle


The friction encountered by a mass sliding on a thin granular layer is
important in a variety of contexts such as walking on sand, braking on
a pebble strewn road, and jamming of joints and bearings in a dusty
environment. Such systems consists of two linearly sheared surfaces 
with a granular layer in between. Assuming for simplicity that the
material properties of the grains and the surfaces are the same, a
basic issue one would like to understand is how the granular case
differs from when solid surfaces slide past each other.  In particular
one would like to know the magnitude of the friction as a function of
layer thickness and the roughness of the boundaries.

A number of studies have examined shear of deep granular layers with linear, couette, and drag
systems~\cite{thompson91,midi04,nasuno97,losert00,mair02,coste04,mueth00,albert00,geng05}.
A shear zone confined over a few grain diameters near one of the
boundaries is usually observed. Qualitative difference between
granular and solid-on-solid stick-slip friction have been noted due to
dilatancy~\cite{bamberger94,nasuno97,hayakawa99,lacome00}. The
friction coefficient and the dilatancy of the shearing layer has been
found to be independent of the shearing rates for deep
layers~\cite{coste04}. Nonetheless, measurements which span the range
from solid-on-solid to granular friction have not been accomplished
and analyzed in any detail. In experiments with granular flows down
rough planes~\cite{pouliquen99}, the inclination required to have
steady motion is observed to decrease with an increase in the layer
thickness, and thus friction may be interpreted as decreasing with
increasing granular layer thickness. However, the material is free to
expand at the top surface which is a crucial difference.

\begin{figure}
  \includegraphics[width=0.9\linewidth]{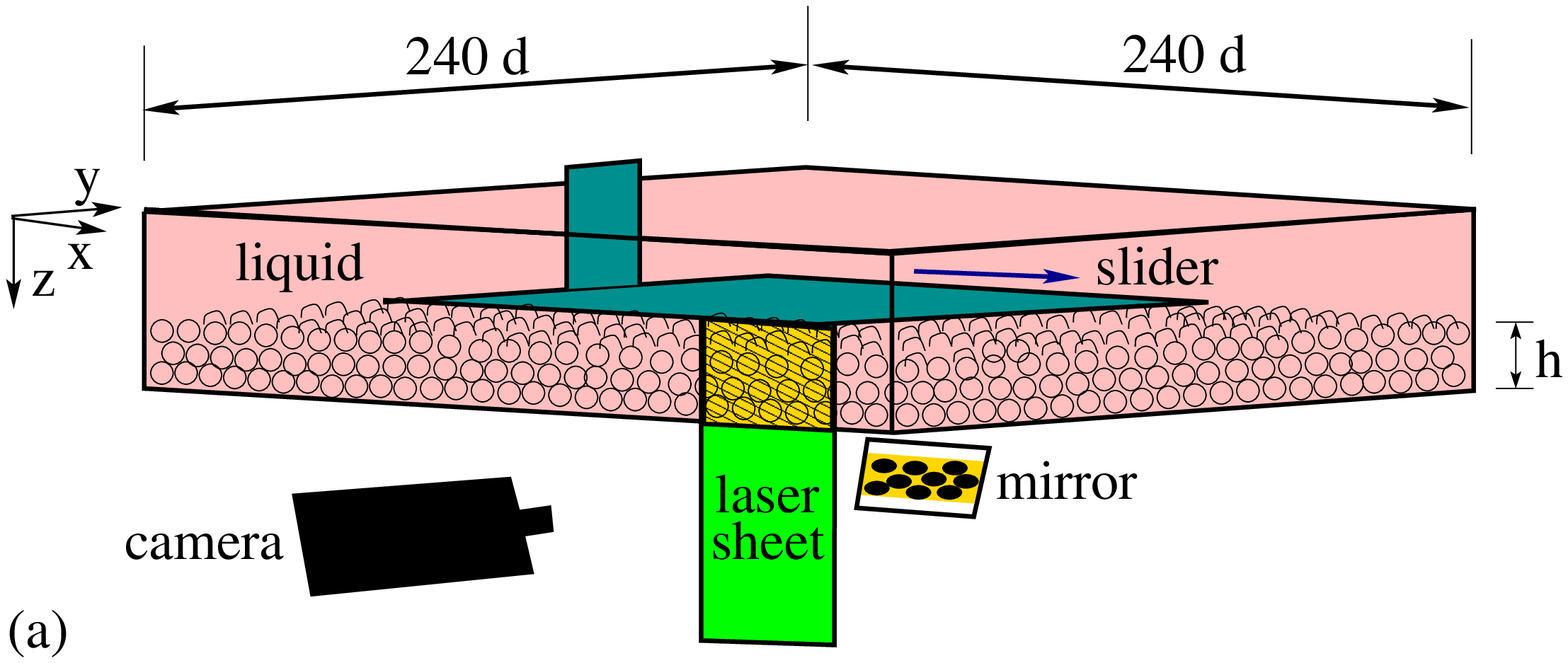}
  \includegraphics[width=0.39\linewidth]{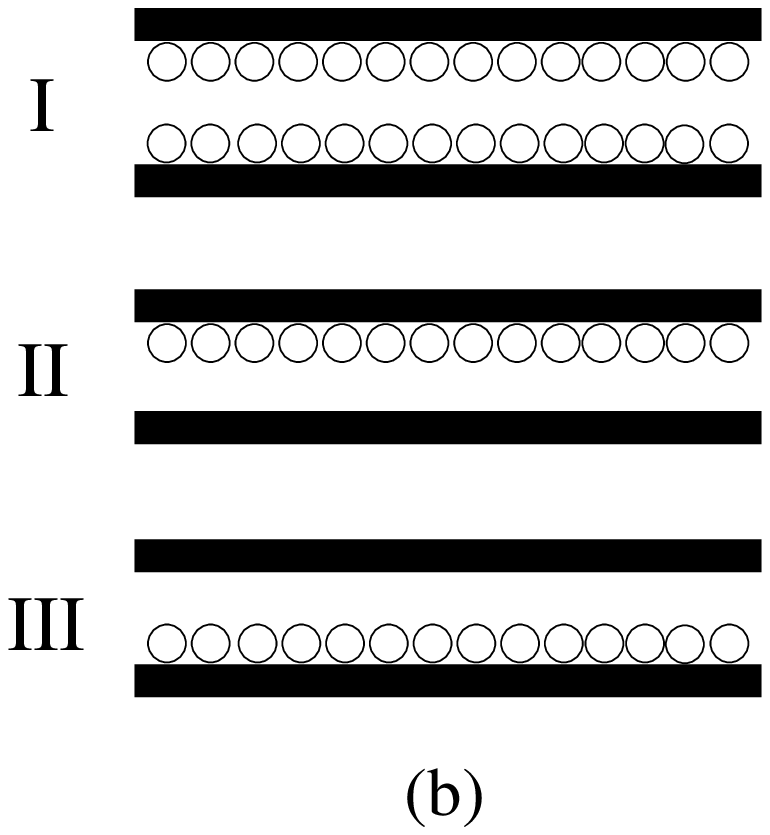}\includegraphics[width=0.46\linewidth]{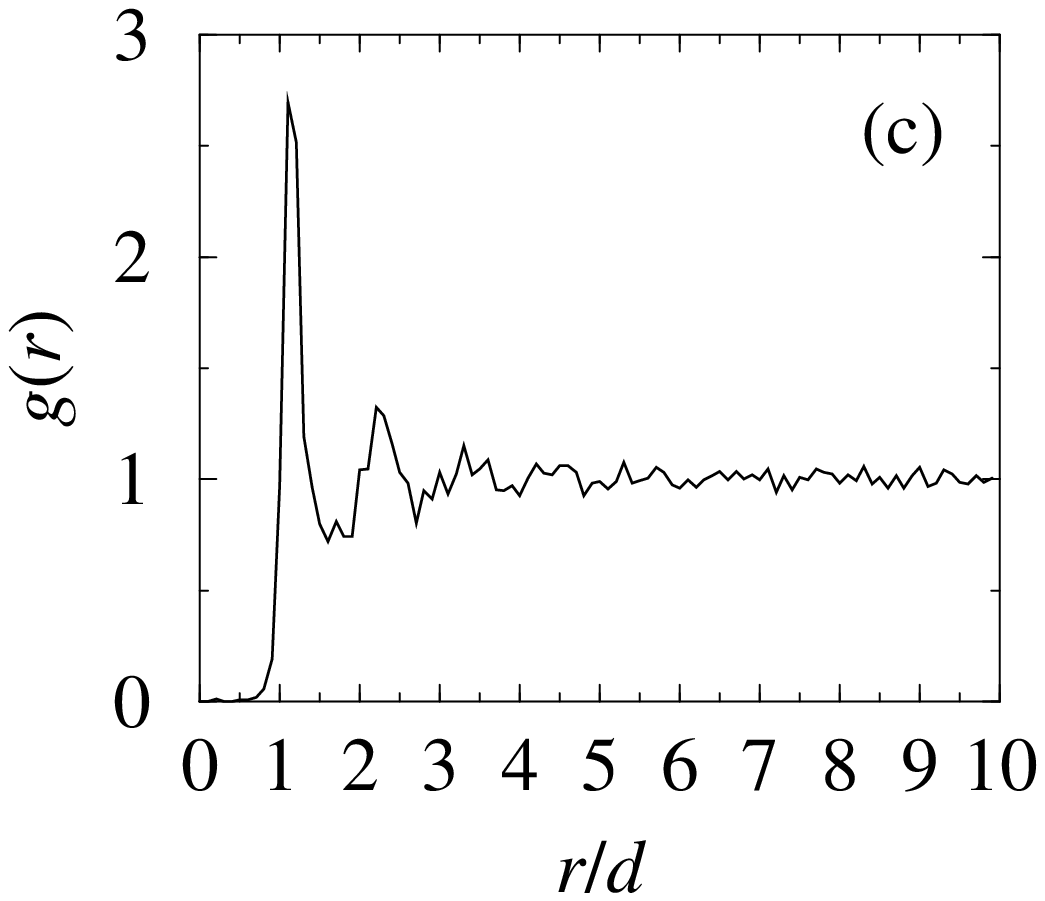}
\caption{(Color Online) (a) Schematic of experimental apparatus. (b) Three kinds
  of boundary surface conditions I: rough-on-rough, II:
  rough-on-smooth, and III: smooth-on-rough. Rough surfaces are
  fabricated by gluing a layer of glass beads to the surfaces. (c)
  Two-point cross correlation function $g(r)$ as a function of
  distance of separation $r$ of two beads on the surface. Peaks at
  multiples of bead diameter are observed indicating absence of
  hexagonal packing.}
\label{apparatus}
\end{figure}

Here, we report the sliding friction of a mass on a thin granular bed
to address open questions in confined and sheared granular matter. 
The friction of the slider with a rough surface moving on a rough
substrate decreases sharply as a grain layer is added, before
increasing and saturating as the bed thickness is increased over ten
layers. Exploiting the fact that the behavior remains unchanged at low shear rates in the presence of an interstitial liquid, we use an index-matching technique to probe the motion
of the grains inside the bed. From these measurements, we propose that the change in the
friction with layer thickness is because
of the increased confinement and locking of a grain relative to its neighbors. 


A schematic of the apparatus is shown in Fig.~\ref{apparatus}(a). A 
rectangular slider of size 100 mm $\times$ 140 mm and mass $m = 0.146$ 
kg is pushed over a granular layer with a linear translating stage
connected to a stepper motor and is similar in design to 
that in Ref.~\cite{nasuno97}. The sliding plate is free to move 
vertically, and the experiment is carried out at 
constant pressure given by the weight divided by surface area.  
A stiff spring with spring constant $k =
1.62 \times 10^4$ N\,m$^{-1}$ is used to couple the slider to the
translating stage and measure the force required to move the slider
with the help of a capacitance displacement sensor.  All the grains (of diameter $d = 1.0 \pm 0.1$ mm) and surfaces used to measure the frictional properties are composed of
soda lime glass. Three combinations of smooth and rough
boundary conditions are used as illustrated in
Fig.~\ref{apparatus}(b). An optically polished glass surface is 
used for the smooth case. A layer of beads is glued on the planar
slider and substrate surfaces in order to obtain rough boundary
conditions. The positions of the grains obtained from an image of the surface, and then characterized by the two-point cross-correlation function $g(r)$ [see
Fig.~\ref{apparatus}(c)] shows no hexagonal order.  

The granular bed with a height $h$ and surface area of 240 mm $\times$
240 mm is prepared by pouring and leveling the grains with a
knife edge. In order to obtain consistent initial conditions for the
granular bed, we first place the slider on the granular bed and push
the slider over a distance of approximately 15$d$ to pre-shear the
system. We then hold the slider for 5 seconds to have a well 
defined pre-aging condition for the contact surfaces. We 
then push the slider with various speeds
$v_{p}$ over a distance of 15$d$ to obtain the spring displacement
with a sampling rate of 1 kHz. The slider either performs stick-slip
or continuous motion depending on the pushing speed and the ratio of
$k$ and $m$~\cite{nasuno97}. For simplicity, we focus on the 
continuous sliding regime. The effective coefficient of sliding 
friction $\mu_{eff}$ is obtained 
by averaging the displacement measured over time and multiplying it 
with $k$ and dividing by the slider weight.

\begin{figure}
  \includegraphics[width=0.65\linewidth]{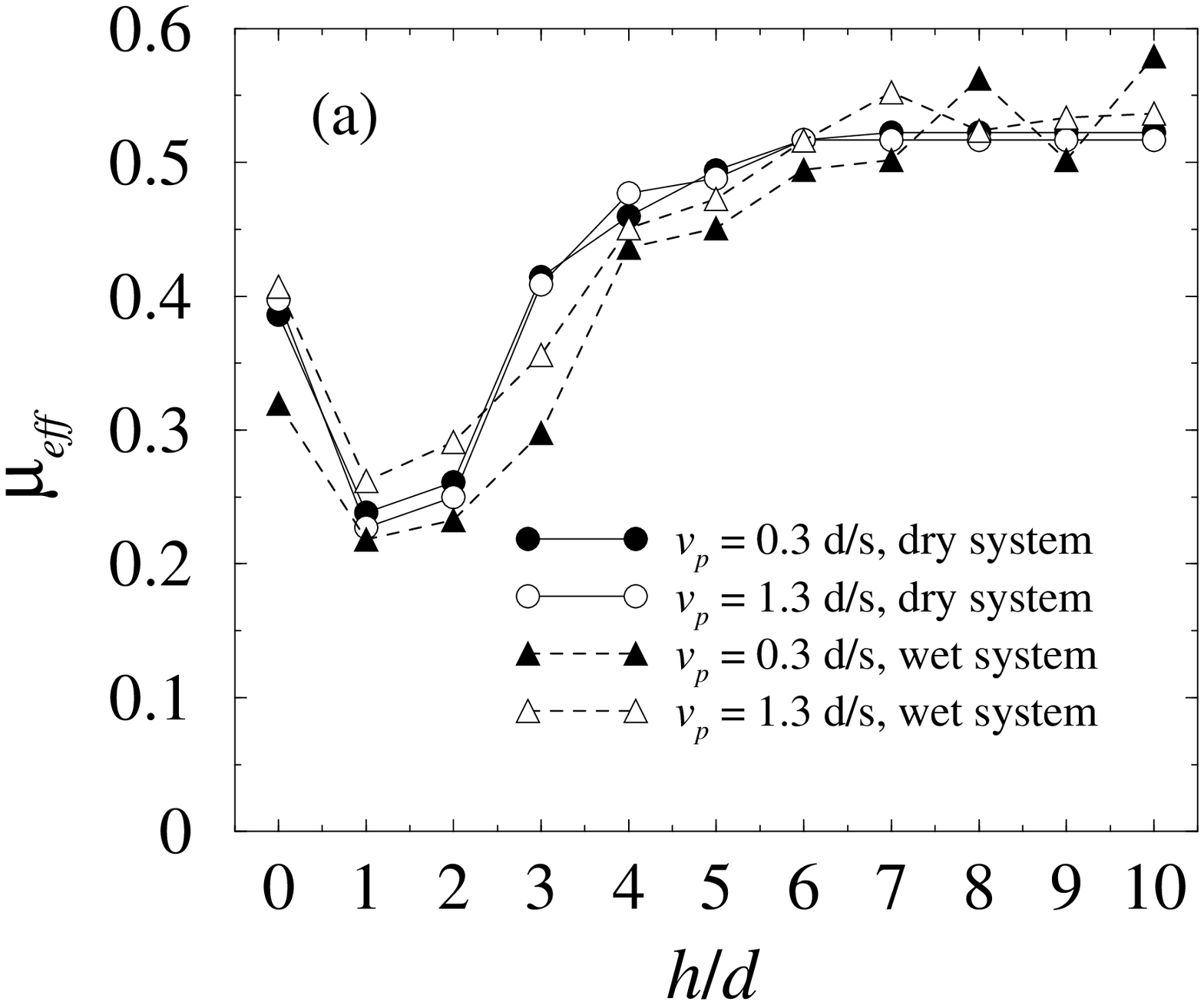}
  \includegraphics[width=0.65\linewidth]{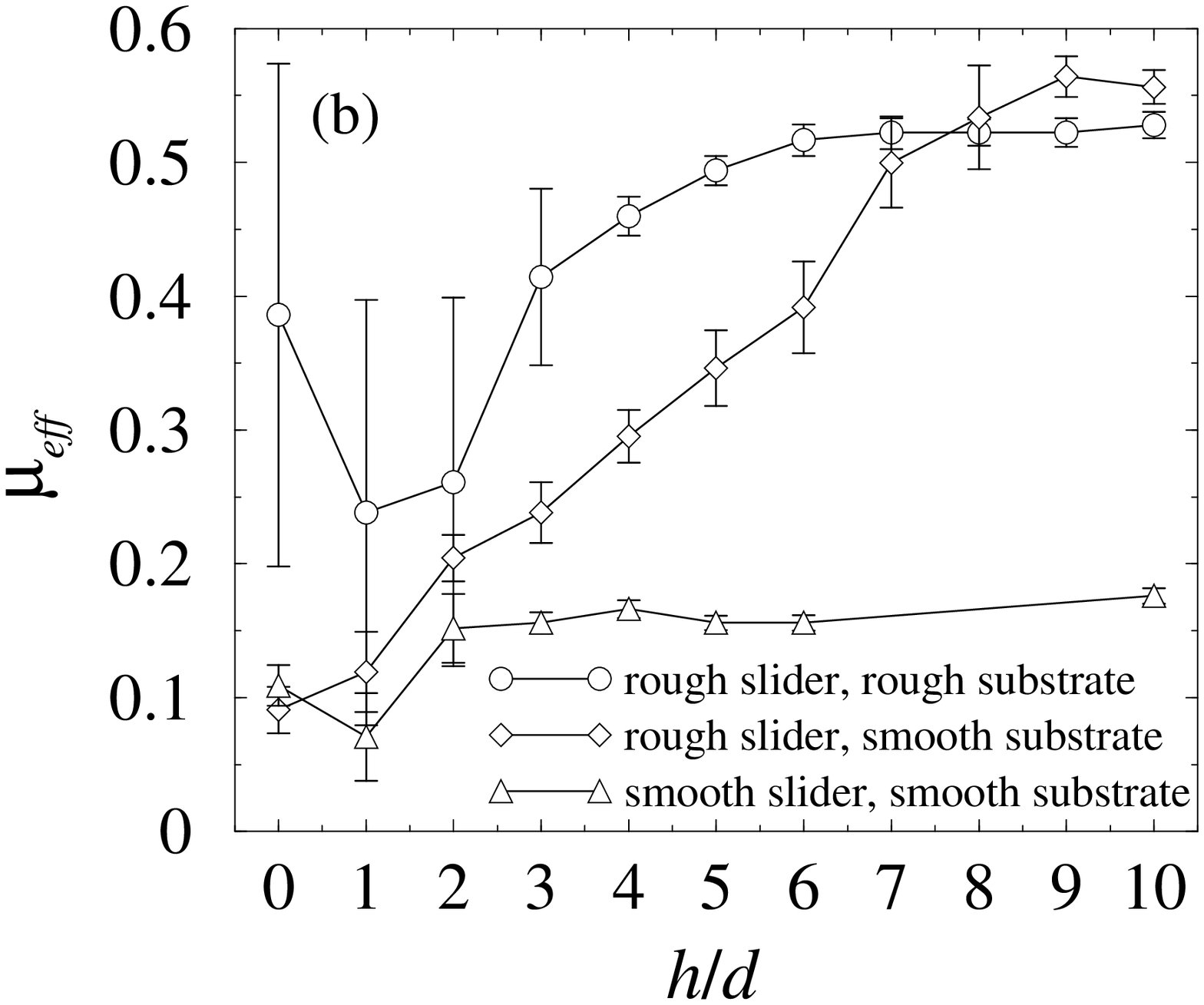}
\caption{(a) Effective coefficient of sliding friction ($\mu_{eff}$) as a function of granular layer of thickness $h$ normalized by grain diameter $d$. All surfaces are rough. (b)
  $\mu_{eff}$ for the case of dry system and three
   boundary conditions shown in Fig.~\ref{apparatus}(b). The measured fluctuations in friction are also shown. ($v_{p} = 0.3 d/$s)}
\label{fig2}

\end{figure}

Figure~\ref{fig2}(a) shows the measured $\mu_{eff}$ for a slider 
with a rough boundary conditions as a function of $h/d$. A 
non-monotonic thickness dependence is observed with
$\mu_{eff}$ first decreasing rapidly as a layer of grains is added
between the slider and the substrate. Then, $\mu_{eff}$ increases and
saturates as the number of layers are increased to about 10.  While
one can imagine that friction between surfaces may decrease if grains
are added due to a lubrication-like effect, the subsequent increase in
friction with $h$ alerts us to the subtlety of the
problem. 

To understand the role of the boundary surfaces, $\mu_{eff}$ 
for three different surface conditions [illustrated in
Fig.~\ref{apparatus}(b)] are plotted in Fig.~\ref{fig2}(b).
First, we focus on the case where the boundaries directly slide
against each other ($h/d = 0$). As long as one of the surfaces is
smooth, the measured value of $\mu_{eff}$ is significantly lower than
when both surfaces are rough. The value depends somewhat on the
solvents used to clean the surfaces and relative humidity. Now the
higher value for the rough surfaces can be understood by considering
the geometry of the rough surfaces. 

For simplicity consider that the rough surfaces can be represented by a row of beads
next to each other as in Fig.~\ref{apparatus}(b). Then, depending on
the relative angle $\theta$ from the vertical where grains on the two
surfaces make contact, the effective friction given by the ratio of
the force required to see continuous motion and the weight of the
slider can be shown to be $\tan(\theta_{0} + \theta)$,
where $\mu_{0} ( = \tan\theta_{0})$ is
the coefficient of friction for two smooth glass surfaces sliding against
each other. Now, $\theta$ can vary between at least 0 and $\pi/6$ depending on 
where neighboring beads touch each other, and therefore using the measured
$\mu_{eff}$ for rough on smooth case for $\mu_{0}$ and using the
average of the angles of the contact, one obtains the effective
friction as 0.5 which is close but somewhat higher than what we
measure for the rough on rough case. It is possible that a closer
match may be obtained by using the actual distribution of contact angles.

Now let us examine the friction dependence on layer thickness. For
a thick or deep enough granular layer, $\mu_{eff}$ depends on the
roughness of the sliding surface and does not depend on the nature of
the substrate. As the number of layers is decreased, $\mu_{eff}$
decreases except when the sliding surface is smooth in which case the
friction encountered remains small and more or less constant 
[see Fig.~\ref{fig2}(b)].

\begin{figure}
  {\includegraphics[width=7cm]{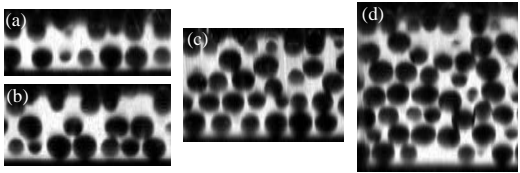}}
\caption{Images of vertical slice of the granular bed away from the side boundaries for
  different layer thicknesses with rough-on-smooth boundary conditions. (a) $h/d = 1$, (b) $h/d = 2$, (c) $h/d  = 4$, and (d) $h/d = 7$.}
\label{image}
\end{figure}

To obtain further insight into this problem, an understanding of the
grain packing and velocity profiles inside the granular layer is
necessary.  A schematic of the setup used for the fluorescent imaging
is also shown in Fig.~\ref{apparatus}(a). The grains are completely
immersed in a liquid with a matching refractive index ($\approx
1.52$). A fluorescent dye 
 with excitation and emission frequencies centered at 525.5 nm and 565 nm,
respectively, is added to the liquid.  As illustrated in
Fig.~\ref{apparatus}(a), planes of the granular bed far from the side
boundaries are illuminated using a 15 mW laser and cylindrical lens
system and imaged with a digital camera. The internal imaging
technique used is similar to that used in
Ref.~\cite{tsai03}. Typical images after rescaling and smoothing are
shown in Fig.~\ref{image} for a few $h/d$. The
apparent size of the beads depends on the distance of the bead center
from the illumination plane. A centroid algorithm is used to find the
particle position to sub-pixel resolution. Imaging at 30
frames per second is sufficient to track the particles and obtain mean
velocities to within 5\%.

Figure~\ref{fig2}(a) also shows $\mu_{eff}$ with the
granular bed immersed in the liquid used for internal imaging. The 
obtained values are observed to be close to those for the dry case 
after correcting for the buoyant force due to the liquid displaced 
by the slider. Thus the measured values do not vary significantly 
at low shear rates. (However, the measured values depart
systematically from the dry case if the slider speed is increased by an order
of magnitude.)

\begin{figure}
  \includegraphics[width=0.85\linewidth]{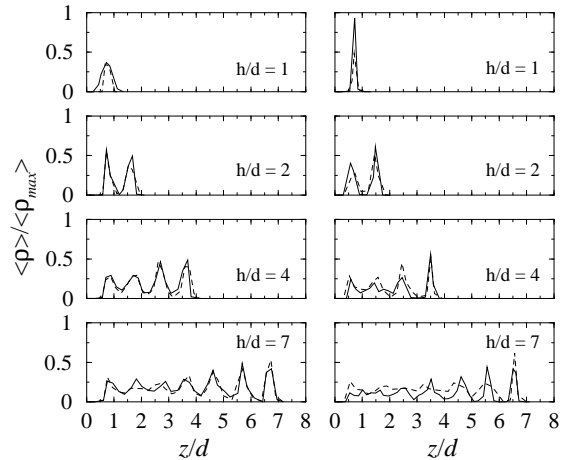}
\caption{(Color Online) Normalized density of the grains as a function of depth
  for the rough-on-rough (first column) and the rough-on-smooth (second column) boundary conditions and slider velocities $v_{p} = 0.3 d/$s (solid lines) and $v_{p} = 1.3 d/$s 
(dashed lines).}
\label{density}
\end{figure}

\begin{figure}
  \includegraphics[width=0.85\linewidth]{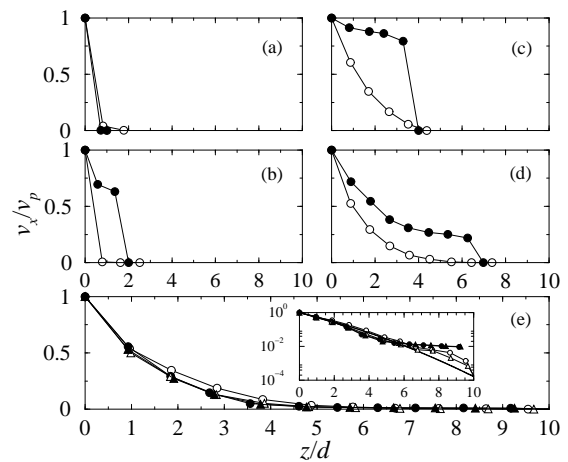}
\caption{Normalized mean velocity of the grains as a function of
  depth for different boundary conditions and slider velocities.
  (a)-(d) $h/d = $1, 2, 4, and 7 respectively and $v_{p} = 0.3 d/s$.
  (e) $h/d = $10, $v_{p} = 0.3 d/s$ ($\circ$), $v_{p} = 1.3 d/s$
  ({\tiny$\triangle$}).  Rough-on-rough case (open symbols) and 
rough-on-smooth case (filled symbols). Inset: Corresponding plot 
in log-linear scale. The thick solid line represents the fit (see text).}
\label{velocity}
\end{figure}

Figure~\ref{density} shows the mean density normalized by the 
maximum packing density $<\rho>/<\rho_{max}>$ of the grains 
as a function of depth $z$ inside the bed for various layer 
thicknesses. Here, $z = 0$ is taken to be the averaged 
lowermost points of the particles glued to the slider.
Peaks are observed which get smaller and broader with increase in height. Thus significant layering is seen especially for lower heights
independent of the overall thickness of the bed.

The corresponding mean velocity normalized by $v_p$ are
plotted as a function of depth in Fig.~\ref{velocity}. For one and for
two layers, the slip almost entirely occurs between the slider and the
granular layer. But as the granular layer thickness is increased, the
slip region grows wider before saturating as the number of layers
approaches 10. As can be noted from Fig.~\ref{velocity}(e), the
velocity profiles are more or less independent of the nature of the
substrate for $h/d \sim 10$. We have fitted the asymptotic velocity
profile with the following
fit: $v/v_{p} = \exp(- a (h/d) - b (h/d)^2)$ where $a = 0.6$, and $b =
0.03$. Thus the form is mostly exponential with a small correction 
similar to that obtained by Mueth
{\em et al}~\cite{mueth00} in a couette geometry far away from side
walls. A somewhat similar profile comprising of a linear part near the shearing surface
followed by a slow exponential decay was obtained theoretically and
numerically for 2D couette flow by Volfson, {\em et al}~\cite{volf04}. 

\begin{figure}
  \includegraphics[width=0.8\linewidth]{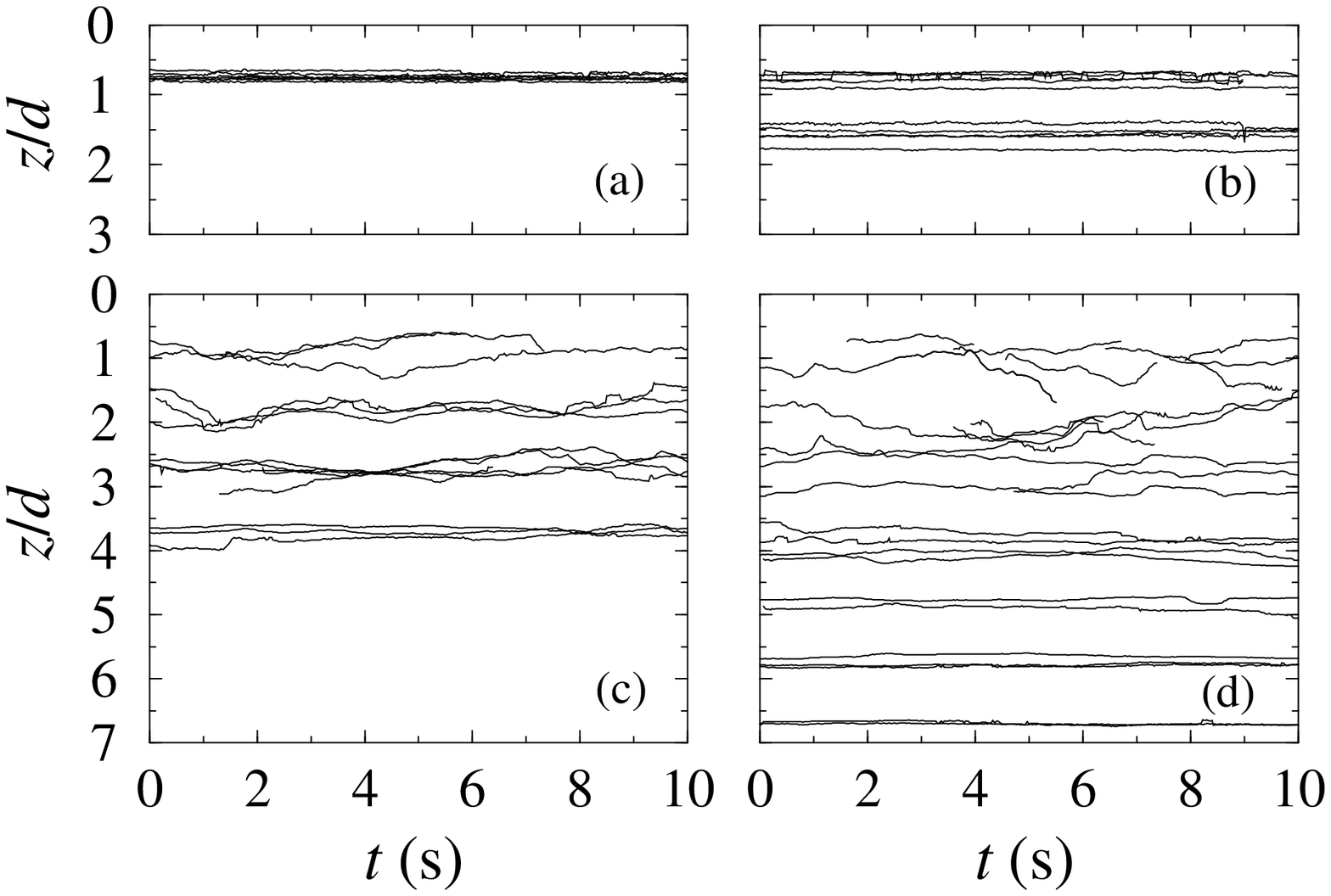}
  \includegraphics[width=0.6\linewidth]{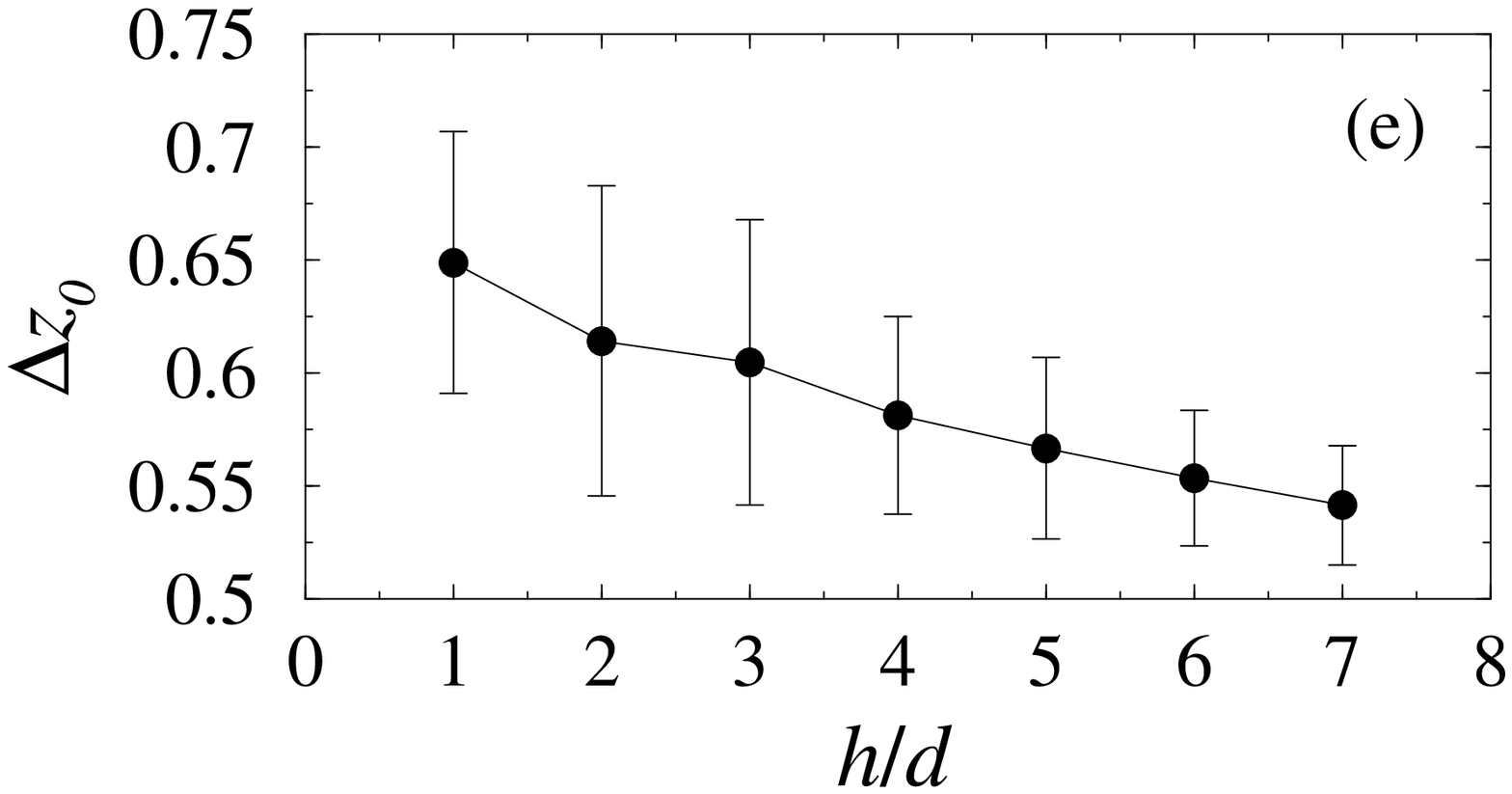}
\caption{Vertical component of the trajectory of a particle for
  different layer thicknesses and $v_{p} = 0.3 d/s$. (a) $h/d = 1$,
  (b) $h/d = 2$, (c) $h/d = 4$, and (d) $h/d = 7$. Rough-on-rough 
  boundary conditions are used. (e) Average gap
  between sliding surface and top layer. The data is averaged over
  all boundary conditions and slider speeds.}
\label{traj}
\end{figure}

Having characterized the overall structure and velocity profiles, we
next plot the vertical component of the trajectories of sample
particles in the granular bed for various $h$ in order to
understand the friction properties. As shown in Fig.~\ref{traj}(a,b),
for $h/d =1$ and 2, the grains more or less remain at the same
height, and do not exhibit significant mean drift [as can be noted
from the velocity profiles plotted in Fig.~\ref{velocity}(a,b)]. From
the movies of the grain motion~\cite{movies}, it may be easily noted 
that the particles more or less fluctuate in the same position. As $h$
is increased and the region of shear increases, grains can be
seen to show increased motion in the vertical direction in addition to
the translation motion along the direction of shear [see
Fig.~\ref{traj}(c,d) and~\cite{movies}].  Because of the
mobility of the grains in the shear zones, the layers rearrange so
that the gap between the shearing surface and the granular layer below
it decreases with $h$. To quantify this trend, we have
measured the gap distance $\Delta z_0$ defined by the distance of the top layer of the particles in the bed from $z = 0$ averaged over all images. The result is plotted in Fig.~\ref{traj}(e).

From these observations one can construct an argument for the increase
and saturation of $\mu_{eff}$ with $h/d$. Because of the
greater gap between the shearing surface and the grains for small $h/d$, it can be noted that the grains are less confined. This allows grains to move around the bumps on the sliding surface more easily,  which lowers the $\mu_{eff}$. However, for greater $h/d$, the grains move until they are jammed against each other and the sliding surface, resulting in lower $\Delta z_0$. This causes the actual sliding surfaces between grains to be at angles other than normal to the horizontal, which results in greater $\mu_{eff}$ due to the additional contribution of the applied force to the normal force between sliding surfaces. Indeed, the measured value of $\mu_{eff}$ for larger $h/d$ is similar to that discussed earlier for rough surfaces directly sliding past against each other.

Our explanation is also consistent with why an increase in friction is not observed in
flows down inclined planes~\cite{midi04} because, a top confining
surface does not exists in that case. While additional support for our argument could be given by examining particle-particle correlation functions within the layers, this is
beyond the capability of our current technique. 

In summary, we have examined the granular layer thickness dependence on the friction encountered by a mass sliding on a granular surface. The observed friction depends on the roughness of the sliding surface and for thin layers it depends on the roughness of the substrate as well. Friction is observed to increase with layer thickness. With the help of grain position data acquired using an index-matching technique, we give an explanation of the increase in friction in terms of the confinement and locking of the grains against its neighbours and the sliding surface. We have also shown how the shear profile changes with layer thickness and surface roughness. In these experiments, gravity clearly breaks the up-down symmetry. It would be  interesting to consider how the phenomena will differ when this symmetry is not broken and will be the subject of future work. 

\begin{acknowledgments}
  We thank I. Nagle, R. O'Donnell and A. Samadani for help with
  acquiring preliminary data and construction of the apparatus. The
  work was supported by the National Science Foundation under Grant Nos.
  DMR-9983659 and CTS-0334587, and a DOE-GLUE grant and the GLUE program of the Department of
  Energy.
\end{acknowledgments}


\end{document}